\documentclass[twocolumn,showpacs,aps,prl,superscriptaddress]{revtex4}
\usepackage{graphicx}
\usepackage{dcolumn}
\usepackage{amsmath}
\usepackage{epsfig}
\input babarsym.tex

\newcommand{\BABARPubYear}    {06}
\newcommand{\BABARPubNumber}  {041}
\newcommand{\SLACPubNumber} {11964}

\def\figurebox#1#2#3{
    \def\arg{#3}
    \ifx\arg\empty
    {\hfill\vbox{\hsize#2\hrule\hbox to #2{\vrule\hfill\vbox to #1{\hsize#2\vfill}\vrule}\hrule}\hfill}
    \else
    {\hfill\epsfbox{#3}\hfill}
    \fi}

\begin{document}
\begin{flushleft}
\babar-PUB-\BABARPubYear/\BABARPubNumber\\
SLAC-PUB-\SLACPubNumber\\
\end{flushleft}

\title{\large \bf Search for \boldmath$\Bp\rightarrow X(3872)\Kp,
X(3872)\rightarrow J/\psi \gamma$ }

%
\author{B.~Aubert}
\author{R.~Barate}
\author{M.~Bona}
\author{D.~Boutigny}
\author{F.~Couderc}
\author{Y.~Karyotakis}
\author{J.~P.~Lees}
\author{V.~Poireau}
\author{V.~Tisserand}
\author{A.~Zghiche}
\affiliation{Laboratoire de Physique des Particules, F-74941 Annecy-le-Vieux, France }
% ...
\collaboration{The \babar\ Collaboration}
\noaffiliation

\begin{abstract}
\noindent In a study of $\Bp\rightarrow \jpsi\gamma\Kp$ decays, we
find evidence for the radiative decay $X(3872)\rightarrow J/\psi
\gamma$ with a statistical significance of $3.4\sigma$. We measure
the product of branching fractions $\mathcal{B}(\Bp\rightarrow
X(3872) \Kp)\cdot\mathcal{B}(X(3872)\rightarrow J/\psi\gamma) =
(3.3\pm1.0\pm0.3)\times10^{-6}$, where the uncertainties are
statistical and systematic, respectively. We also measure the
branching fraction $\mathcal{B}(\Bp\rightarrow \chi_{c1}
\Kp)=(4.9\pm0.2\pm0.4)\times10^{-4}$. These results are obtained
from $(287\pm3)$ million \BB decays collected at the \FourS
resonance with the \babar\ detector at the PEP-II $B$ Factory at
SLAC.
\end{abstract}


\maketitle

\def\bfDbar    {\kern 0.2em\overline{\kern -0.2em D}{}\xspace}
\def\bfDzb     {\ensuremath{\bfDbar^0}\xspace}
\def\bfDstarz  {\ensuremath{D^{*0}}\xspace}
\def\bfDzDzb   {\ensuremath{\bfDzb {\kern -0.16em \bfDstarz}}\xspace}

\noindent The $X(3872)$ state was discovered by the Belle
Collaboration \cite{belle_1} in the decay $\Bp\rightarrow
X(3872)\Kp$ \cite{charge_conjugation}. This signal was confirmed by
the \babar\ Collaboration \cite{babar_1}, as well as the CDF and
D$\O$ Collaborations \cite{fermilab_1}. Interpreting this new state
has been challenging. Its narrow width, mass near the \bfDzDzb
threshold, and small branching fraction for the radiative decay
$X(3872)\rightarrow\chicone\gamma$ have made it difficult to
identify the $X(3872)$ with any of the predicted charmonium states
\cite{charmon1}. Alternate proposals have been made, including a
\bfDzDzb molecule \cite{molecule}, or a diquark-antidiquark state
\cite{diqdiq}. Evidence for the radiative
$X(3872)\rightarrow\jpsi\gamma$ decay in $\Bp\rightarrow X(3872)\Kp$
would determine the C-parity of the $X(3872)$ state to be positive,
limiting the conventional charmonium assignment options while
remaining consistent with \bfDzDzb molecule model predictions.

A number of other new states have recently been found. The Belle
Collaboration has claimed the discovery of a broad resonance in $B$
decays, referred to here as the $Y(3940)$ state \cite{belle_y3940}.
The nature of this state is unknown, and there is no reason to yet
preclude $\Bp\rightarrow Y(3940)\Kp, Y(3940)\rightarrow\jpsi\gamma$
as a possible decay channel. Belle has also identified a possible
$\chi_{c2}'$ charmonium candidate in two photon production, referred
to here as the $Z(3930)$ state \cite{belle_z3930}. This state could
be produced in $B$ decays, and if the tentative $\chi_{c2}'$
charmonium assignment holds true, it should decay radiatively to
$\jpsi\gamma$ (albeit at a rate predicted \cite{highercharmonia} to
be quite small).

We study the decay chain $\Bp\rightarrow\ccbar\Kp$, where \ccbar
decays radiatively to $\jpsi\gamma$, and the \jpsi subsequently
decays to a lepton pair. The notation \ccbar represents conventional
charmonium, such as the triplet $\chi_{cJ}(1P)$ states, or any state
with positive C-parity for which the $\jpsi\gamma$ decay is not
forbidden.

The data sample for this analysis consists of $(287\pm3)$ million
\BB pairs collected with the \babar\ detector at the \pep2
asymmetric \epem collider. This represents 260\invfb of data taken
at the \FourS resonance. The \babar\ detector is described in detail
elsewhere \cite{babar_nim}. The innermost component of the detector
is a double-sided five-layer silicon vertex tracker (SVT) for
precise reconstruction of \B-decay vertices. A 40-layer drift
chamber (DCH) measures charged-particle momentum. A ring-imaging
detector of internally reflected Cherenkov radiation (DIRC) is used
for particle identification. Energy deposited by electrons and
photons is measured by a CsI(Tl) crystal electromagnetic calorimeter
(EMC). These detector subsystems are surrounded by a solenoid
producing a 1.5-T magnetic field. The flux return for the magnet is
instrumented with a muon detection system composed of resistive
plate chambers (RPC). For the most recent 51\invfb of data, a
portion of the muon system has been replaced by limited streamer
tubes (LST) \cite{lst}.

A $\jpsi\rightarrow\ellell$ candidate is reconstructed by combining
a pair of oppositely charged muon or electron candidates having an
invariant mass compatible with the nominal \jpsi\ mass. An attempt
is made to recover energy loss due to bremsstrahlung by searching
for photons near electron candidates. Candidates for \jpsi are then
combined with a candidate kaon and a photon to form a
$\Bp\rightarrow\jpsi\gamma\Kp$ candidate.

The $\jpsi\rightarrow\epem$ candidates are formed with electrons
(and bremsstrahlung photons) with $2.950 < m(\epem(\g)) < 3.170$
\gevcc. Candidates for $\jpsi\rightarrow\mumu$ require muons with
$3.060 < m(\mumu) < 3.135$ \gevcc. The \ccbar candidate is
reconstructed from the mass-constrained \jpsi and a photon with
$E_{\g}$ greater than 30 \mev. Additional selection criteria are
applied to the shape of the lateral distribution ($LAT$)
\cite{photon_lat} and azimuthal asymmetry (measured by the Zernike
moment, $A_{42}$) \cite{photon_a42} of the photon-shower energy
deposited in the EMC. The radiative $\gamma$ candidate is rejected
if, when combined with any other $\gamma$ from the event, the
invariant mass is consistent with the $\pi^{0}$ mass (see Table
\ref{tab_cuts}). The ratio of the second and zeroth Fox-Wolfram
moments $(R_{2})$ \cite{fox_wolfram} is used to separate isotropic
\Bp events from typically anisotropic continuum background events.
The mass of the \ccbar candidates, $m_{\ccbar}$, is calculated by
constraining the \Bp candidate to the nominal \Bp mass.

To identify \B candidates, we use two kinematic variables, $m_{B}$
and $m_{miss}$. The unconstrained mass of the reconstructed \B
candidate $m_{B} = \sqrt{E_{B}^{2}/c^{4}-p_{B}^{2}}$, where $E_{B}$
and $p_{B}$ are obtained by summing the energies and momentum of the
particles in the candidate $B$ meson, respectively. The missing mass
is defined through $m_{miss}=\sqrt{(p_{\epem}-\hat{p}_{B})^{2}}$,
where $p_{\epem}$ is the four-momentum of the beam \epem system and
$\hat{p}_{B}$ is the four-momentum of the \B candidate after
applying a \Bp mass constraint. These variables are uncorrelated by
construction, and are advantageous for analyzing $B$ decays in which
a particle in the final state has poorly measured energy. Events
with a correctly reconstructed \Bp decay should have values equal to
the nominal \Bp mass for both kinematic variables.

To best separate signal from background, the signal selection
criteria are chosen based on Monte Carlo (MC) samples by maximizing
the figure of merit $n_{S}/(\alpha/2+\sqrt{n_{B}})$ \cite{punzi}
where $n_{S}$ and $n_{B}$ are numbers of signal and background
events, respectively, and $\alpha$ represents the minimum level of
significance desired. For this analysis, $\alpha=3$ is chosen. The
optimization is performed by varying the selection values for
$m_{\epem(\gamma)}$, $m_{\mumu}$, $R_{2}$, photon $LAT$, photon
$A_{42}$, and the photon $\pi^{0}$ veto, while requiring $m_{B}$ and
$m_{miss}$ to be within $100$\mevcc of the nominal \Bp meson mass.
The optimized criteria used in this analysis are summarized in Table
\ref{tab_cuts}.

\begin{table}[t]
\begin{center}
\caption {Summary of acceptance criteria for candidate events.}
\begin{tabular}{c}
\hline \hline
Region\\
\hline
$2.950<m_{\epem(\gamma)}<3.170$ GeV/c$^{2}$ \\
$3.060<m_{\mumu}<3.135$ GeV/c$^{2}$ \\
$R_{2}<0.35$ \\
$LAT<0.5$ \\
$A_{42}<0.1$ \\
Reject $122<m_{\pi^{0}\rightarrow\gamma\gamma}<145$ \mevcc\\
\hline \hline
\end{tabular}
\label{tab_cuts}
\end{center}
\end{table}

We extract the signal with an unbinned two-dimensional extended
maximum-likelihood (ML) fit to the kinematic variables $m_{B}$ and
$m_{miss}$ in $10\mevcc$ bins of $m_{\ccbar}$. Fits failing to
converge or lacking statistics are combined with adjacent
$m_{\ccbar}$ bins to ensure fit success. The probability density
functions (PDFs) for signal extraction are the product of
independent fits in $m_{\B}$ and $m_{miss}$, defined separately for
signal and background events.

The signal PDFs are determined from Monte Carlo simulation of
$\Bp\rightarrow\chicone\Kp$ and $\Bp\rightarrow X(3872)\Kp$ decays.
The $m_{\B}$ and $m_{miss}$ distributions of
$\Bp\rightarrow\ccbar\Kp$ signal events are both modeled by a
functional form similar to a Gaussian with asymmetric tails, $f(x)=$
exp$\left[-(x-m)^{2} /
\left(2\sigma^{2}_{\pm}+\alpha_{\pm}(x-m)^{2}\right)\right]$, where
the $\pm$ subscript indicates different parameter values on either
side of the central peak. The signal PDFs for these two \ccbar modes
are found to be equivalent to one another within statistical
uncertainty.

The background consists of two parts, a combinatoric component with
a flat distribution in the kinematic variables $m_{B}$ and
$m_{miss}$, and a component that peaks in $m_{miss}$ composed of \B
decays similar to our decay mode. The peaking background events are
mostly from $\Bp\rightarrow J/\psi\Kp \pi^{0}, \pi^{0}\rightarrow
\gamma \gamma$ and $\Bp\rightarrow J/\psi K^{*+}, K^{*+}\rightarrow
\Kp \pi^{0}, \pi^{0}\rightarrow \gamma\gamma$ decays. These events
are incorrectly reconstructed as the desired final state if one of
the photons from the $\pi^{0}$ decay is undetected. This background
does not peak in the other kinematic variable $m_{B}$, nor in
$m_{\ccbar}$. The only doubly-peaking background may arise from
$\Bp\rightarrow J\psi K^{*+}, K^{*+}\rightarrow \Kp \gamma$. These
events can peak in both $m_{B}$ and $m_{miss}$, but the branching
fraction for this decay mode is small and can be ignored. The
simulation also indicates that the combinatorial background is
almost entirely due to $B$ decays.

The background PDFs are fitted to events from generic \BpBm, \BzBzb,
\qqbar ($q=u,d,s,c$), and \tautau MC samples. In $m_{B}$, all
background events are modeled by the tail of a wide Gaussian
function. The $m_{miss}$ distribution of background events is
parametrized by the ARGUS background shape \cite{argus} for the
combinatoric component, while the peaking component is characterized
by a Gaussian function.

The maximum-likelihood fit returns the number of
$\Bp\rightarrow\ccbar\Kp$ signal events, $N_{sig}$, in each 10
\mevcc $m_{\ccbar}$ bin. The number of signal events is found by
fitting to the $N_{sig}$ versus $m_{\ccbar}$ results with functions
representing the \ccbar mass distribution of each signal mode. Based
on Monte Carlo simulation of the $\chi_{cJ=0,1,2}$ and $X(3872)$
\cite{xwidth} decays, the $m_{\ccbar}$ shape for each of these
signals is individually parameterized with a double Gaussian
distribution. In the fit to the ML results, the Gaussian means,
widths, and ratios of the areas are fixed to the values determined
from the MC simulation, with the heights of the peaks permitted to
float. As $N_{sig}$ can also include non-signal events peaking in
both $m_{B}$ and $m_{miss}$, a first order polynomial in $N_{sig}$
versus $m_{\ccbar}$ was included to account for the level
doubly-peaking backgrounds. The number of \ccbar events is
calculated from the area of the fitted Gaussians above this
background.

The effectiveness of the signal extraction method is validated on
Monte Carlo samples for $\chi_{c0,1}$ and $X(3872)$. It is found
that the proximity of the large $\chi_{c1}$ signal peak introduces a
significant bias in the measurement of a $\chi_{c2}$ signal with
this method. Therefore we do not quote results for the \chictwo
mode. Successful performance of the $X(3872)$ extraction is verified
on Monte Carlo generated samples for numbers of events similar to
the measured value, as well as for the case of a null result.

The efficiency is determined by calculating the fraction of the
events generated in Monte Carlo simulation that survived the
analysis selection criteria from Table \ref{tab_cuts} and are
returned by the fitting procedure. Standard \babar\ corrections are
applied to account for particle identification and tracking
differences found between simulation and data. These corrections are
at the level of a few percent. The resulting efficiencies are
$(16.8\pm0.2)\%$ for the $X(3872)$ mode, $(13.3\pm0.2)\%$ for
\chiczero and $(13.5\pm0.3)\%$ for \chicone, where the errors are
statistical.

Systematic uncertainties on the branching fractions are reported in
Table \ref{tab_systematics}. Sources include uncertainty in the
number of \BB pairs, uncertainty in the secondary branching
fractions for $\chi_{c0,1}\rightarrow \jpsi\gamma$ and
$\jpsi\rightarrow\ellell$, PDF parameterization uncertainty due to
MC statistics, uncertainty in the $m_{\ccbar}$ parameterization,
particle identification, tracking and photon corrections, effects
due to fit technique (such as choice of $m_{\ccbar}$ bin width and
fit starting point), and uncertainty in the true $X(3872)$ mass and
width. The uncertainties due to MC statistics, $m_{\ccbar}$ and
$X(3872)$ mass were evaluated by varying the individual parameter
values by $1\sigma$ from their measured values and finding their
effect on the signal yield. The largest source of uncertainty (aside
from secondary branching fractions which are beyond the control of
this analysis) is due to the variation in signal yield with the
choice of PDF parameter values. In the case of the $X(3872)$ signal,
the total uncertainty is dominated by statistical rather than
systematic errors.

\begin{table}[t]
\begin{center}
\caption {Summary of systematic uncertainties. The uncertainty due
to secondary branching fractions (BFs) does not apply to the product
of branching fraction results.} \label{tab_systematics}
\begin{tabular}{lccc}
\hline \hline Source & $\chi_{c0} (\%)$ & $\chi_{c1} (\%)$ & $X(3872) (\%)$\\
\hline $B$ counting & $1.1$ & $1.1$ & $1.1$\\
Secondary BFs & $8.5$ & $5.4$ & $1.0$ \\
MC statistics & $16.5$ & $3.2$ & $8.7$\\
$m_{\ccbar}$ shape & $3.1$ & $1.3$ & $1.5$\\
Particle ID & $2.0$ & $2.0$ & $2.0$\\
Tracking & $3.6$ & $3.6$ & $3.6$\\
Photons & $1.8$ & $1.8$ & $1.8$\\
Fit technique & $1.7$ & $1.7$ & $1.7$\\
$X(3872)$ mass/width & - & - & $2.0$\\
\hline Total & $19.5$ & $8.1$ & $10.3$\\
\hline \hline
\end{tabular}
\end{center}
\end{table}

\begin{figure}
\begin{center}
\epsfig{file={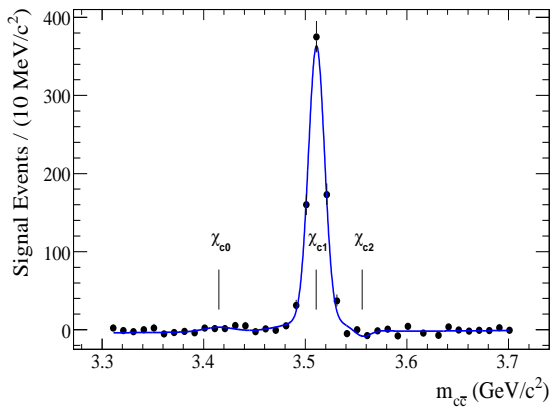},width={3.0in},height={2.25in}}
\caption{Number of extracted signal events versus $m_{\ccbar}$ for
$\chi_{cJ}$. The solid curve is the fit to the data. The $\chi^{2}$
per degree of freedom for this fit is 48.7/34.} \label{fig_chic}
\end{center}
\end{figure}

Figure \ref{fig_chic} shows the fit to $m_{\ccbar}$ in the mass
range $3.311<m_{\ccbar}<3.711$ \gevcc. We extract $27.9\pm11.7$
\chiczero events and $807.2\pm33.3$ \chicone events. Using our
signal extraction efficiencies, we calculate the product of
branching fractions
$\BR(\Bp\rightarrow\chicone\Kp)\cdot\BR(\chicone\rightarrow\jpsi\gamma)=(1.76\pm0.07\pm0.12)\times10^{-4}$
and
$\BR(\Bp\rightarrow\chiczero\Kp)\cdot\BR(\chiczero\rightarrow\jpsi\gamma)=(6.1\pm2.6\pm1.1)\times10^{-6}$,
where the first error is statistical and the second is systematic.
Taking the branching fractions for
$\chi_{c0,1}\rightarrow\jpsi\gamma$ from \cite{pdg}, we calculate
$\BR(\Bp\rightarrow\chicone\Kp)=(4.9\pm0.2\pm0.4)\times10^{-4}$ ,
and $\BR(\Bp\rightarrow\chiczero\Kp)=(4.7\pm2.0\pm0.9)\times10^{-4}$
corresponding to the $90\%$ confidence level upper limit of
$\BR(\Bp\rightarrow\chiczero\Kp)<7.5\times10^{-4}$. These branching
fraction results are consistent with the current world average
\cite{pdg}, and in the case of $\Bp\rightarrow\chicone\Kp$, more
precise.

\begin{figure}
\begin{center}
\epsfig{file={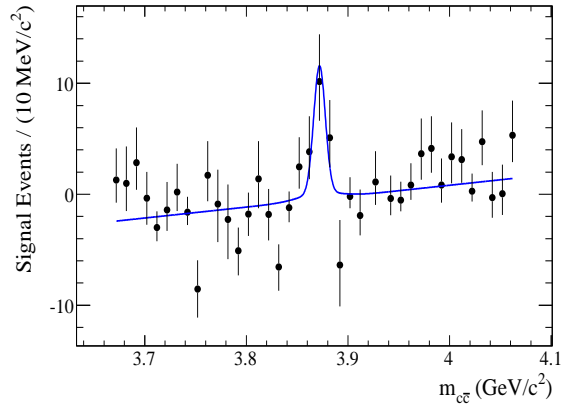},width={3.0in},height={2.25in}}
\caption{Number of extracted signal events versus $m_{\ccbar}$ for
the $X(3872)$ region. The solid curve is the fit to the data. The
$\chi^{2}$ per degree of freedom for this fit is 57.8/37.}
\label{fig_x3872}
\end{center}
\end{figure}

We extract the number of $X(3872)$ signal events in the mass range
$3.672<m_{\ccbar}<4.072$ \gevcc and find $19.2\pm5.7$ events (Fig.
\ref{fig_x3872}). We derive the product of branching fractions
$\BR(\B\rightarrow X(3872)
\Kp)\cdot\BR(X(3872)\rightarrow\jpsi\gamma) =
(3.3\pm1.0\pm0.3)\times10^{-6}$. The statistical significance of
this signal, taken to be the square root of the difference in
$\chi^{2}$ values between the fit in Fig. \ref{fig_x3872} and a
similar fit assuming zero signal events, is $3.4\sigma$.

Additional fits are performed to search for the $Y(3940)$ and
$Z(3930)$ states by adding a resonance in the appropriate mass
region. The measurement of the $Y(3940)$ state from
\cite{belle_y3940} finds a mass of $3943\pm17$ \mevcc and width of
$87\pm34$ \mevcc, while the $Z(3930)$ state is found to have a mass
of $3929\pm5$ \mevcc and width of $29\pm10$ \mevcc
\cite{belle_z3930}, where the statistical and systematic
uncertainties have been combined in quadrature. We model the mass
resolution for the decays of each of these states to $\jpsi\gamma$
by a Gaussian function in $m_{\ccbar}$ with the mean and sigma fixed
to the Belle measurements. Because the masses and photon energies
are similar, we assume the same efficiency for these modes as for
the $X(3872)$ state. We find $-16\pm34$ events and $-5.4\pm8.3$
events for the $Y(3940)$ and $Z(3930)$ states, respectively. We
define an upper limit on the product of branching fractions by
assuming a Gaussian distribution for the number of signal events and
its uncertainty, and integrate over the physically-allowed region
from 0 to 90\% of the total area around the mean. Systematic errors
are estimated from the contributions listed for the $X(3872)$ in
Table \ref{tab_systematics}. Uncertainties on the $Y(3940)$ and
$Z(3930)$ masses and widths dominate entirely. The total systematic
uncertainty on the product of branching fractions is 101\% for the
$Y(3940)$ and 22\% for the $Z(3930)$. To account for the width
uncertainty, it was varied by $1\sigma$ from the measured value and
the largest resulting upper limit retained. Using these basic
assumptions, we calculate $\BR(\B\rightarrow Y(3940)
\Kp)\cdot\BR(Y(3940)\rightarrow\jpsi\gamma) < 1.4\times10^{-5}$ and
$\BR(\B\rightarrow
Z(3930)\Kp)\cdot\BR(Z(3930)\rightarrow\jpsi\gamma) <
2.5\times10^{-6}$ at the 90\% confidence level.

In summary, we measure the branching fraction
$\BR(\Bp\rightarrow\chicone\Kp)=(4.9\pm0.2\pm0.4)\times10^{-4}$ and
determine a $90\%$ confidence level upper limit of
$\BR(\Bp\rightarrow\chiczero\Kp)<7.5\times10^{-4}$. We find the
product of branching fractions $\BR(\B\rightarrow X(3872)
\Kp)\cdot\BR(X(3872)\rightarrow\jpsi\gamma) =
(3.3\pm1.0\pm0.3)\times10^{-6}$, with a statistical significance of
$3.4\sigma$. This provides evidence of the radiative decay
$X(3872)\rightarrow J/\psi \gamma$ and of charge parity $C=+$ for
the $X(3872)$ state. We search for radiative decays of the Y(3940)
and Z(3930) states to $\jpsi\gamma$ in the $\Bp\rightarrow\ccbar\Kp$
channel, and find no evidence for such modes.

We are grateful for the excellent luminosity and machine conditions
provided by our \pep2\ colleagues, 
and for the substantial dedicated effort from
the computing organizations that support \babar.
The collaborating institutions wish to thank 
SLAC for its support and kind hospitality. 
This work is supported by
DOE
and NSF (USA),
NSERC (Canada),
IHEP (China),
CEA and
CNRS-IN2P3
(France),
BMBF and DFG
(Germany),
INFN (Italy),
FOM (The Netherlands),
NFR (Norway),
MIST (Russia),
MEC (Spain), and
PPARC (United Kingdom). 
Individuals have received support from the
Marie Curie EIF (European Union) and
the A.~P.~Sloan Foundation.


\begin{thebibliography}{99}
\bibitem{xwidth}
The $X(3872)$ state is generated with a natural width of zero in the
Monte Carlo simulation.
\end{thebibliography}
\end{document}